\documentclass[letter,scriptaddress,twocolumn, prl,showkeys]{revtex4}
	\usepackage{amsmath}
	\usepackage{makeidx}
	\usepackage{amsfonts}
	\usepackage[ansinew]{inputenc}
	\usepackage[usenames,dvipsnames]{pstricks}
	\usepackage{epsfig}
	\usepackage{pst-grad} 
	\usepackage{pst-plot} 
	\usepackage[colorlinks,hyperindex]{hyperref}
	\hypersetup
	{
		colorlinks,%
		citecolor=black,%
		linkcolor=black,%
		urlcolor=black,%
	}

\makeindex

\begin{document}

\title { Heterogeneous Dynamical Relaxations in a Lane Forming Liquid}

\author{Suman Dutta}
\email{Email:sumand@bose.res.in}
\affiliation{Department of Chemical Biological and Macromolecular Sciences\\
S. N. Bose National Centre for Basic Sciences \\ Block-JD, Sector-III, Salt Lake, Kolkata 700098 \\ India.\\
}

\date{23 November, 2016 }

\begin{abstract}
The connection between domain relaxations at individual scales and the collective heterogeneous response in non-equilibrium systems is a topic of profound interest in recent times. In a model system of constantly driven oppositely charged binary colloidal suspension, we probe such relaxations as the elongated lanes of likely charges interact with increasing field in the orthogonal plane using
Brownian Dynamics simulations. We show that the system undergoes a structural and dynamical cross-over: from an initial fast-relaxing homogeneous phase to a heterogeneous lane phase following
a slow relaxation via an intermediate phase with mixed relaxation. Our observations also affirm the coexistence of multiple time and length scales in this phase. Unlike glasses, these time and length
scales are not separated by orders of magnitude, they are comparable in magnitudes. Using a phenomenological model, we, further, reinforce that these competing relaxations are a manifestation
of heterogeneity in diffusion which is rather generic in systems with such competing relaxations. \\
\end{abstract}

\maketitle

{\it Introduction: }Driven away from equilibrium, soft materials exhibit fascinating phenomena, ranging from dynamic patterns to active self-assembly, and often generates technological applications in diverse areas of science \cite{soft1}. Yet the underlying microscopic description is mostly unexplored, for the particle dynamics depends explicitly on the structural changes and external field competes with inter-particle interactions\cite{soft2}. This interplay between structure and dynamics governs the heterogeneous transport processes in complex media and hence develops widespread interests in a variety of inter-disciplinary subjects bridging physics, chemistry, biology and engineering\cite{soft1,soft2,rev1}. 

Laning \cite{rev1,rev2,lane1,lane2,lane3,lane3a,netz, lane4,lane5,lane6,lane7,lane8,lane9, lane10,lane11} is one such example realized in a host of simple systems like army ants\cite{ant}, pedestrian movements\cite{ped}, granular media\cite{granule}, dusty plasma\cite{plasma}, dipolar microswimmers \cite{swimmers} and often considered as a generic model of non-equilibrium systems where two species of particles are driven against each other \cite{ped,granule,ant,plasma,lane1,lane2,lane3,lane3a,netz, lane4,lane5,lane6,lane7,lane8,lane9,lane10,lane11}. 
Colloids mimic the phenomenon: applying a constant electric field, the system of binary charges crosses over from a homogeneous mixture to a state with columnar lanes of likely charged particles elongated parallel to the field\cite{soft1,soft2,rev1,rev2,lane1,lane2,lane3,lane3a, netz, lane4,lane5,lane6,lane7,lane8,lane9,lane10,lane11}. Report shows there is a `locked-in' situation followed by a slow dynamics as the system gets more heterogeneous in the lane state due to the presence of both slow and fast mobile particles leading to a dynamical heterogeneity\cite{lane8} as the primitive lanes grow with field \cite{lane9}. Recently, Dutta {\it et. al.} have shown that the laning occurs via a novel 'pre-lane' phase with anomalous dynamical responses leading to a heterogeneity in diffusion\cite{lane10}, like that in super-cooled liquids\cite{smk}, yet the nature of dynamics differs at a single particle level. Although predicted as a generic feature in soft materials\cite{nat_mat}, heterogeneity in diffusion lacks macroscopic realization and microscopic insights. Furthermore, bridging the gap between heterogeneity in dynamics and structural relaxation would trigger interests in fundamental exploration of complex transport processes of technological importance \cite{appl1,appl2}. Moreover, it aids inputs into the understanding of collective response in a class of non-equilibrium systems with similar inherent structural heterogeneity being influenced by field which needs not to be external, in general. 

In this paper, we explore the steady-state dynamical relaxations of a binary colloidal suspension in the presence of a constant electric field using Brownian Dynamics simulations \cite{erm} in terms of a two and four point dynamic correlation functions. The two-point function, the Overlap function \cite{cd}, $Q(t)$ shows a crossover in dynamics: from an initial faster relaxation in the homogeneous phase to a slow down in the lane phase via a phase with mixed relaxation. Beyond the intervening phase, the in-plane motion slows as the lanes grow stronger with increasing field. The Dynamic susceptibility\cite{smk-cd}, $\chi_{4}(t)$ crosses over from a peaked form in the homogeneous phase to a doubly-split form in the lane phase : The prevalent peak at higher $t$ indicates response due to dominant slow particles while the one at lower $t$ is due to faster relaxation by low populating fast particles. However, in the intermediate phase, $\chi_{4}(t)$ broadens indicating signatures of coexisting length-scales in the system. Lanes grow monotonically as the collective interactions between the lanes in the orthogonal plane increase, though the spread in the distribution of these length-scales is, primarily, a non-monotone. Furthermore, our phenomenological calculations show that these competing relaxations, indeed, manifest a heterogeneous response via heterogeneity in diffusion reiterating Ref. \cite{lane10}, albeit distinct from glassy systems.

{\it Methods: } We take equi-molar binary mixture of positively ($N_{+}$) and negatively ($N_{-}$) charged colloidal particles of diameter $\sigma(=1\mu m$)($N_{+}$=$N_{-}$=2000) in a cubic box of length ($L=21.599\sigma $) dissolved in a medium with viscosity($\eta=0.1cP$) at temperature $T(=298K)$ as in \cite{lane10}. The pair-interactions between two particles located at positions $\vec{r_{i}}$ and $\vec{r_{j}}$, with charges $q_{i}$ and $q_{j}$, respectively is given by $V(r_{ij} )= V_{SC} + V_{Repulsion}$ with  $ V_{SC} = V_{0} [q_{i} q_{j} /(1+\kappa \sigma /2)^{2} ][\exp (-\kappa \sigma ((r_{ij} /\sigma )-1))/r_{ij} /\sigma ] $ and $V_{repulsion} =\varepsilon [(\sigma /r_{ij} )^{12} -(\sigma /r_{ij} )^{6} ]+\frac{1}{4}$ for $r_{ij} <2^{1/6} \sigma $  and zero, elsewhere [9] with $r_{ij} =|\mathop{r_{i}}\limits^{\to } -\mathop{r_{j}}\limits^{\to } |$. Here, $\kappa $ is the inverse screening length, $V_{0 }$ the interaction strength parameter and $\varepsilon =4\left|q\right|^{2} V_{0} (1+\kappa \sigma /2)^{2} $ with $q_{i}=q_{j}=q$ [9]. We use $\kappa \sigma (=5.0)$and $V_{0}^{*} =\left|q\right|^{2} V_{0} /k_{B} T(=50.0)$. The BD simulations \cite{erm} are carried out using discretized form of Langevin's equation in over-damped limit with integration time step $\Delta t=0.00005\tau_{\beta}$. We use $\Gamma (=3\pi \sigma \eta )$, the viscous damping and $\mathop{F_{i}}\limits^{\to} (t)$ the fluctuating force with variance $2D_{0} \delta _{\alpha \beta } \delta _{ij} \delta (t-t')$ where $\alpha $,$\beta $ denote the cartesian components and $D_{0}$ the Einstein-Stokes Diffusion coefficient with $\Gamma D_{0} =k_{B} T$,$k_{B} $ being the Boltzmann constant. The simulations use $\tau _{\beta } =(\sigma^{2} /D_{0} $ ) as unit time, $\sigma$ the length unit and $k_{B} T$ the energy unit. We switch on the electric field $f(=\left|q\right|f_{0} \sigma /k_{B} T)$ once we equilibrate the system with $f=0$ from random configurations for $50\tau_{\beta}$. The field is kept on for $100\tau_{\beta}$ so that for all $f$ (within the observation window), the system reaches steady state where statistics are stored for more $50\tau_{\beta}$. We generate $N_{T}(=20)$ set of parallel trajectories with different initial configurations. During our analysis we average over these $N_{T}$ Brownian trajectories generated using different seeds with different initially equilibrated configurations.

{\it Results and Analysis:} We show the particle configurations in Fig. 1. For $f=0$ we observe a homogeneous mixture of opposite charges (data not shown). The situation remains homogeneous for $f=50$ [Fig. 1(a)]. However, for $f=150$, we find tiny domains of likely charged particles [Fig. 1(b)]. On increasing $f$ further, these domains takes the form of network like structures [Fig. 1(c)] proliferated in z direction as lanes as in previous studies \cite{lane5,lane10}. Though there is a steady flow, the lanes continue to re-order in the transverse plane as seen in different steady state configurations (not shown) for a particular $f$. However, the structural order parameter \cite{lane5} in steady state s does not change with $t$ (Data not shown). To understand the underlying structural rearrangements, we look for the dynamical relaxation of these structures. 

\begin{figure}[h]
\includegraphics[angle=0,scale=0.12]{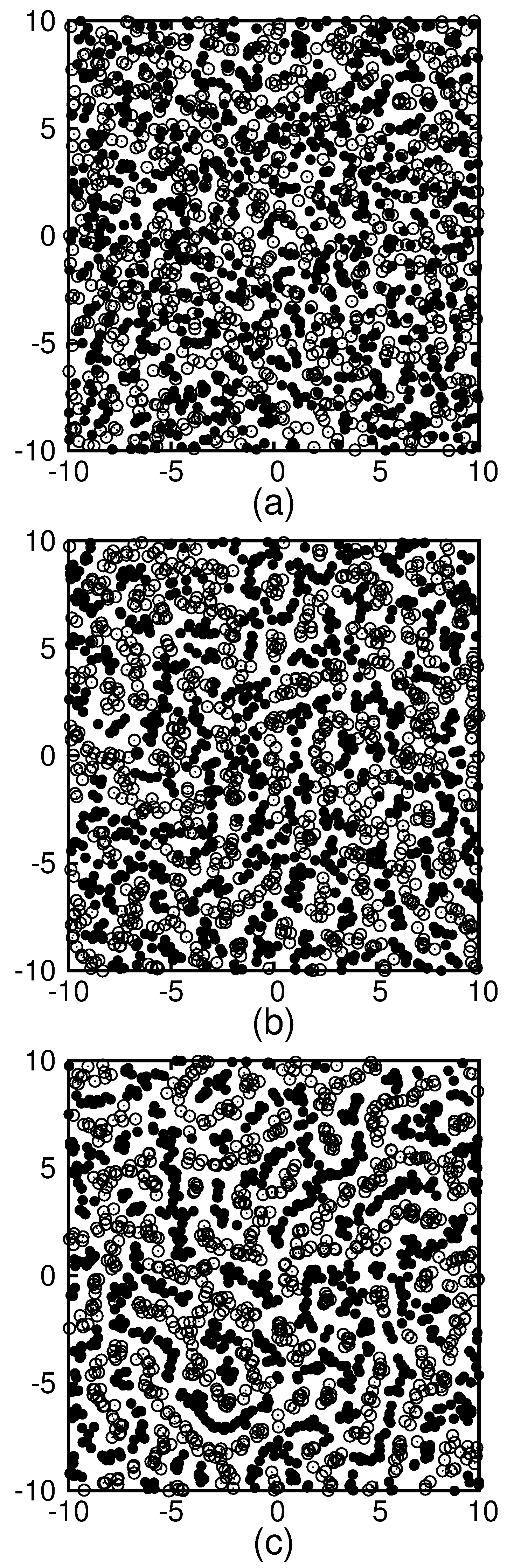}
\caption{Projection of the particles in plane transverse to the drive for (a)$f=50$ (b)$f=150$ (c)$f=300$. Here open circles and filled circles denote $+ve$ and $-ve$ charges respectively.}
\end{figure}

Such structural relaxation is generally interpreted via the self Overlaps, given by $\tilde{q}^{(+)}(t)\sim \frac{1}{N_{+}} \sum_{i=1} ^{N_{+}} \psi(|\vec{r}(t_{0})-\vec{r}(t+t_{0})|)$ \cite{smk-cd}. Here, $\psi(r)=1$ if $r\leq 0.3\sigma$ and $\psi(r)=0$ elsewhere. To compute such overlaps between particle configurations in the orthogonal plane, separated by time $t$, we evaluate $\tilde{q}^{(\pm)}_{(\bot)}(t)$ for both $+ve$ and $-ve$ charges respectively. Both charges behave identically, so we focus on $\tilde{q}^{(+)}_{(\bot)}(t)$. We compute $\tilde{q}^{(+)}(t)$ for several steady state configurations separated by $t$. Thus, there exists a probability distribution function of these $\tilde{q}^{(+)}(t)$, given by $P(\tilde{q}^{(+)}(t))$. We observe $P (\tilde{q}_{(\bot)}^{(+)}(0)) \sim \delta(\tilde{q}_{(\bot)}^{(+)}(0) - N_{+})$ while for $t \neq 0$, $P (\tilde{q}_{(\bot)}^{(+)}(t))$ shifts to lower values of $\tilde{q}_{(\bot)}^{(+)}(t)$ with increasing $t$. In Fig. 2(a) we show $P (\tilde{q}_{(\bot)}^{(+)}(t))$ for various $f$ for one typical $t$(=0.05 $\tau_{\beta}$). For all $f$, $P (\tilde{q}_{(\bot)}^{(+)}(t))$ is Gaussian at low $t$. However, for $f=150$, $P (\tilde{q}_{(\bot)}^{(+)}(t=0.15 \tau_{\beta}))$ deviates from Gaussian for intermediate $t$ while it is again a Gaussian for high $t$ (data not shown). Fig. 2(a) shows $P (\tilde{q}_{(\bot)}^{(+)}(t))$ for different values of $f$ at $t=0.05 \tau_{\beta}$. For $f=50$, $P (\tilde{q}_{(\bot)}^{(+)}(t))$ peaks at $\tilde{q}_{(\bot)}^{(+)}(t)\approx 0.89$. At $t=0.05$, this peak first shifts to lower values of $\tilde{q}_{(\bot)}^{(+)}(t)$ and then again shifts to the higher values of $\tilde{q}_{(\bot)}^{(+)}(t)$. For $f=150$ and $f=300$, we find that the peaks in $P (\tilde{q}_{(\bot)}^{(+)}(t))$ exist at $\tilde{q}_{(\bot)}^{(+)}(t)\approx 0.87$ and $\tilde{q}_{(\bot)}^{(+)}(t)\approx 0.91$ respectively. So the shift of the peak has got some non-monotonic dependence. This behaviour may be associated with the dynamical changes in the system reported in Ref. \cite{lane10}. 

The shift of peaks in $P(\tilde{q}_{(\bot)}^{(+)}(t))$ is given by the Overlap Function, $Q_{(\bot)}^{(+)}(t) (= < \tilde{q}_{(\bot)}^{(+)}(t) > \sim \int \tilde{q}_{(\bot)}^{(+)}(t)P(\tilde{q}_{(\bot)}^{(+)}(t))d \tilde{q}_{(\bot)}^{(+)}(t))$ and is shown in Fig. 2(b). For all $f$, $Q^{(+)}_{(\bot)}(0)=N_{+}$  and they decay monotonically with $t$. However, the decay trend differ with $f$. $f=50$ shows a fast decay in $Q^{(+)}_{(\bot)}(t)$. For $f=150$, the decay is slower than that for $f=50$. Then, $Q^{(+)}_{(\bot)}(t)$ continues to slow down even though $f$ is increased further. For $f=300$, $Q^{(+)}_{(\bot)}(t)\approx 0.25$, twice the value $(\approx 0.12)$ of that for $f=50$ at $t=1\tau_{\beta}$ while we find $Q^{(+)}_{(\bot)}(t=1)\approx 0.18$ for $f=150$. Beyond this, at higher $t(\approx 10\tau_{\beta})$, the decay is significantly slower ($\approx4$ times) compared to that for $f=50$. Clearly, the trend of the long $t$ data with increasing $f$ does not match with the trend with the data in the small $t$ limit. For $f=50$, $Q^{(+)}_{(\bot)}(t)\sim t^{\alpha}$ where $\alpha \approx -0.86$ while we find  $Q^{(+)}_{(\bot)}(t)\sim e^{-t^{- \beta}}$ in the time window $10\tau_{\beta}<t<40\tau_{\beta}$ for $f=150$ with $\beta\approx 0.37$. $Q^{(+)}_{(\bot)}(t)$ again shows a power law dependence with $\alpha\approx -0.52$ for $f=300$ where the fitting time window is limited to $1\tau_{\beta}<t<20\tau_{\beta}$.
 
\begin{figure}[h]
\includegraphics[angle=0,scale=0.12]{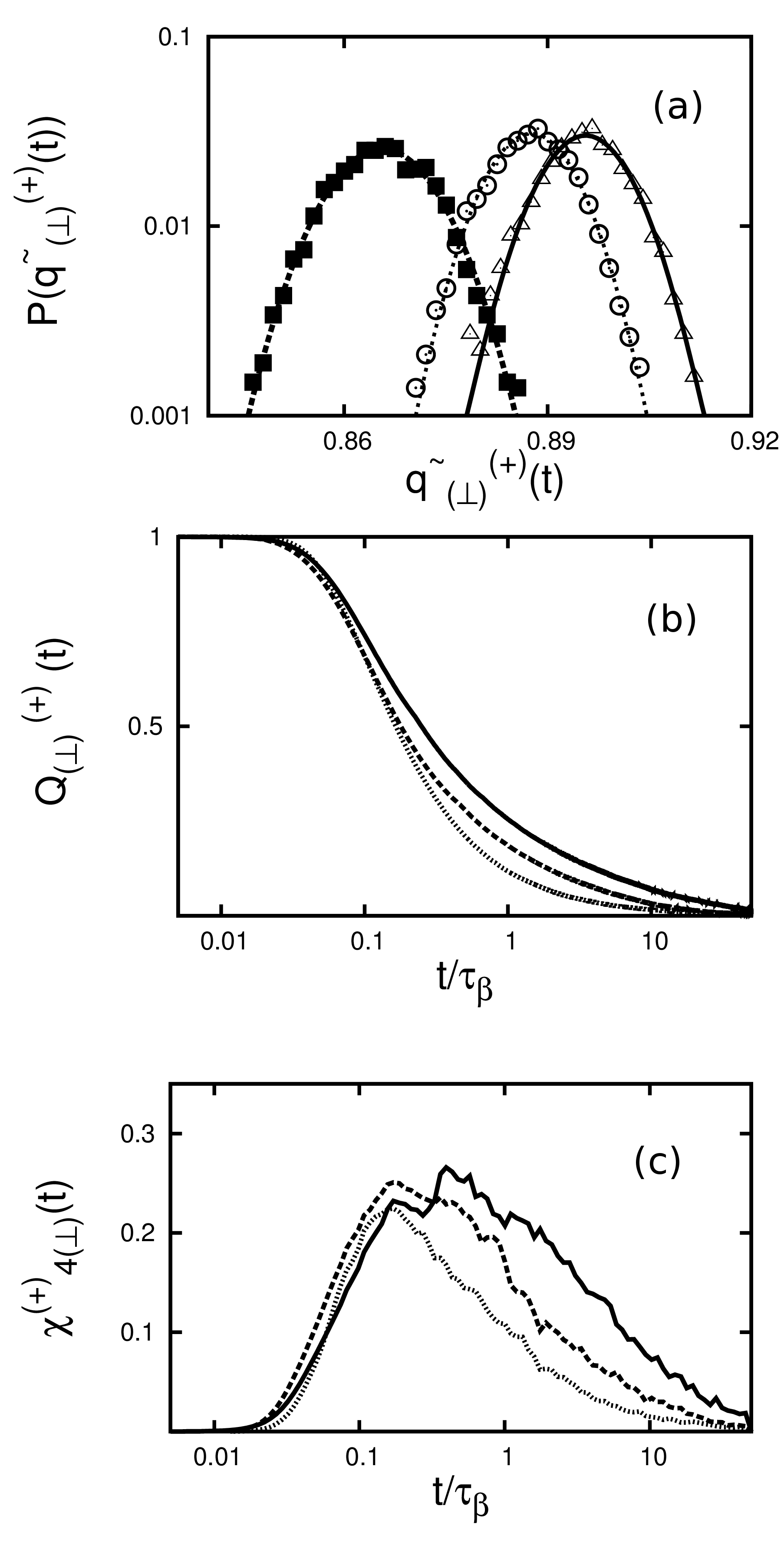}
\caption{(a) Distribution of Overlaps, $P(\tilde{q}_{(\bot)}^{(+)}(t))$ for different $\tilde{q}_{(\bot)}^{(+)}(t)$ is shown for $t=0.05 \tau_{\beta}$: $f=50$ (open circles), $f=150$ (filled squares) and $f=300$(open triangles). Lines show the fitted Gaussian curves. (b) Dependences of $Q_{(\bot)}^{(+)}(t)$ on $t$ for three regimes: fast-segregation $f=50$ (dashed line), mixed relaxation $f=150$ (dotted line) and slow relaxation $f=300$ [Main Panel] (c) Dynamical Susceptibility, $\chi^{(+)}_{4}(t)$ for three regimes: $f=$50 (dotted line),150 (dashed line) and 300 (solid line)}
\end{figure}

$\chi^{(\pm)}_{4}(t)$ is given in terms of the dynamic fluctuations in $Q(t)$, $\chi^{(\pm)}_{4}(t)=<Q^{(\pm)2}(t)>-<Q^{(\pm)}(t)>^{2}$ \cite{cd,smk-cd}. It peaks at $t=\tau_{4}$ , the structural relaxation time \cite{smk-cd}. We show the evolution of $\chi^{(+)}_{4(\bot)}(t)$ with $t$ for different $f$ in Fig. 2(c). $\chi^{(+)}_{4(\bot)}(t)$ for $f=0$ grows with $t$ and shows a peak at $t=\tau_{4}$ [Fig. 2(c)] just as in normal liquid\cite{smk-cd}. Then, for $f=50$, the peak shifts to lower value of $t$ than that for $f=0$ indicating initial phase segregation and hence, the faster relaxation in  $Q^{(+)}_{(\bot)}(t)$. For $f=150$, $\chi^{(+)}_{4(\bot)}(t)$ grows and broadens with no prominent peak, showing the coexisting time-scales of structural relaxation with comparable magnitudes in the system. On increasing $f$ further, $\chi^{(+)}_{4(\bot)}(t)$ shows two distinct peaks for $f=300$. Since the peak in $\chi^{(+)}_{4(\bot)}(t)$ for $f=50$ is due to the relaxation by the faster particles in the system, the predominant peak in $\chi^{(+)}_{4(\bot)}(t)$ at higher $t$ for $f=300$ corresponds to the response by slow particles with larger stake.

Structural heterogeneity induces slowing down of dynamics in a system \cite{td}. Hence, we now look at the effective interactions mediated by the interacting clusters present in the system. The in plane pair distribution functions \cite{allen,ef_int} are given by $g^{(++)}(r_{\bot})$ and $g^{(+-)}(r_{\bot})$ for the like and cross species pairs respectively. We observe $g^{(++)}(r_{\bot})\sim g^{(--)}(r_{\bot})$ and $g^{(+-)}(r_{\bot})\sim g^{(-+)}(r_{\bot})$ (Data not shown). In the transverse plane, the Effective interaction ($V_{eff}^{(++)}(r_{\bot})$) between a pair of $+ve$ particles in presence of other particles is given by the relation: $g^{(++)}(r_{\bot})\sim \exp(-\beta V_{eff}^{(++)}(r_{\bot}))$ leading to $V_{eff}^{(++)}(r_{\bot})\sim -\beta \ln g^{(++)}(r_{\bot})$ \cite{ef_int}. Similarly, we have $V_{eff}^{(+-)}(r_{\bot})\sim -\beta \ln g^{(+-)}(r_{\bot})$. In Fig. 3(a) we show the dependence of $V_{eff}^{(+-)}(r_{\bot})$ (Main Panel) and $V_{eff}^{(++)}(r_{\bot})$ (Inset) on $r_{\bot}$. For $f\neq0$, we observe a peak in $V_{eff}^{(+-)}(r_{\bot})$ and a dip in $V_{eff}^{(++)}(r_{\bot})$ for $r_{\bot}\approx 0$ that grows with increasing $f$. This indicates with increasing $f$, the system experiences growing enhanced effective attraction between like charge-pairs while an increased effective repulsion between cross charge-pairs. The increase in both the effective interactions follow the monotonic growth in steady-state structural order parameter with increasing $f$, reported in earlier studies \cite{lane10,lane5}. Here, $V_{eff}^{(++)}(r_{\bot})$ deviates from $V^{(++)}(r_{\bot})$ for $f\ne 0$ due to increase in collective behavior of a particular species. This excess contribution is due to the increased many body contribution arising out of the competing particle interaction coupled to the field. Hence, for a high $f$, effective interaction between two like charges is more influenced by the neighboring like particles in the same lane while the same for two opposite charges is dominated by the effective interaction of two lanes of the opposite charges. Since $V_{eff}^{(++)}(r_{\bot})\neq V^{(++)}(r_{\bot})$ and $V_{eff}^{(+-)}(r_{\bot})\neq V^{(+-)}(r_{\bot})$ for $f\neq 0$, $V_{eff}^{(++)}(r_{\bot}) \sim V^{(++)}(r_{\bot})+ \delta V^{(++)} (r_{\bot})$ where $\delta V^{(++)} (r_{\bot})$ contains contributions due to direct correlation \cite{ef_int} in particle pairs which, here, is coupled to the applied $f$. For $f=0$, one ends up with $\delta V^{(++)} (r_{\bot})\approx \delta V^{(+-)} (r_{\bot}) \approx 0$ in the low density limit. 

\begin{figure}[h]
\includegraphics[angle=0,scale=0.1]{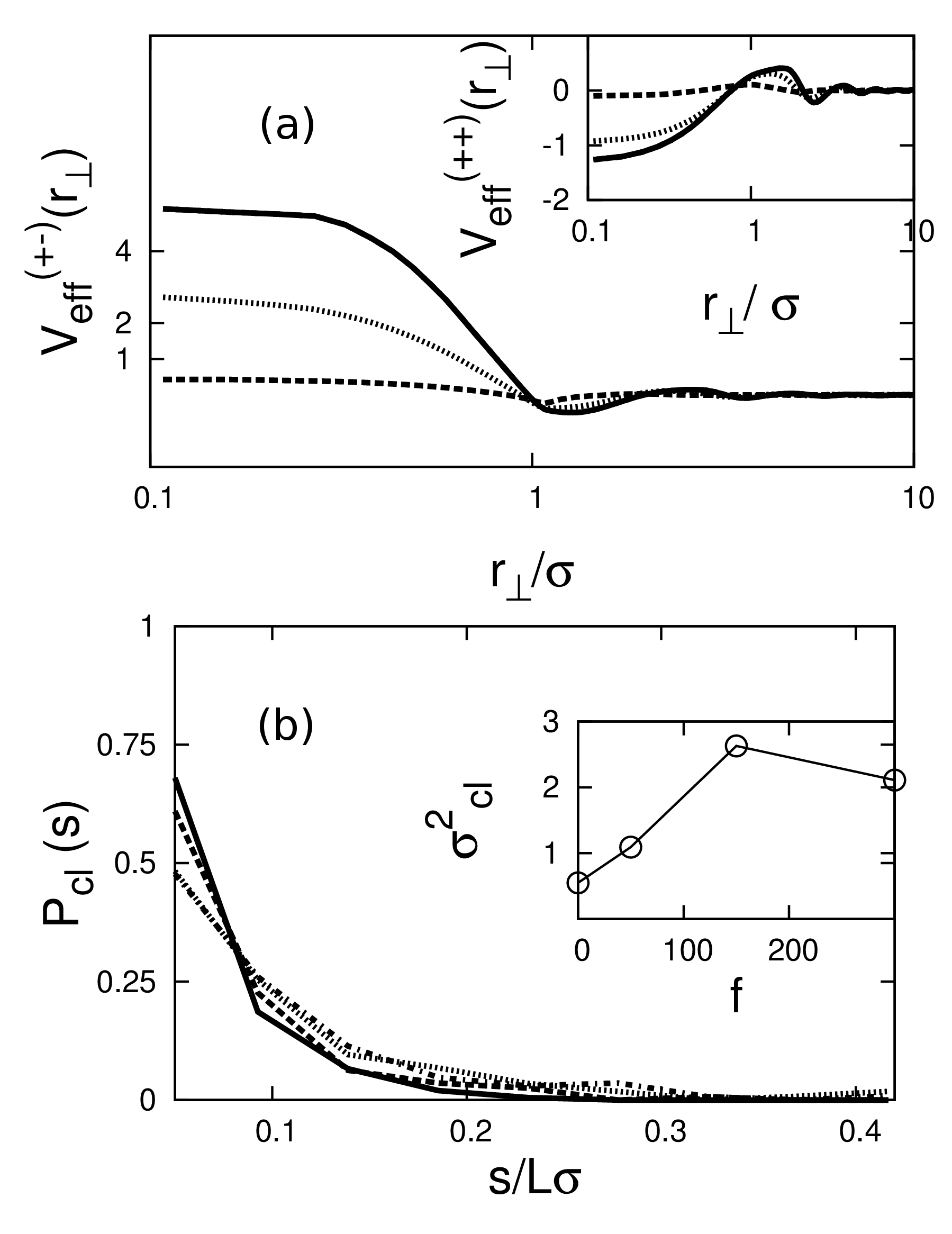}
\caption{(a) Effective Interactions: $V_{eff}^{(+-)}(r_{\bot})$ vs $r_{\bot}$ for $f=$ 50(dashed line),150 (dotted line)and 300 (bold line) Inset. $V_{eff}^{(++)}(r_{\bot})$ vs $r_{\bot}$ for $f=$ 50(dashed line),150 (dotted line) and 300 (bold line)(b) Dependence of $P_{cl}(s)$ on $s/L\sigma$ for $f=$,0(bold line),50(dashed line, 150(dot-dashed line) ,300(dotted line) }
\end{figure}

Motivated by Ref.\cite{lane3a}, in order to find the associated distribution of the length-scales present in the system, we compute the probability of a particle to be a part of a particular cluster of size $s$. In a cluster of likely charged particles, we subsequently add particles of the same species within a critical separation \cite{allen} in three dimensions, $r_{cl}^{(++)}$ (for +ve particles) and $r_{cl}^{(--)}$ (for -ve particles) to obtain the size of the cluster $s$. This is repeated for all the particles of the same charge to obtain the cluster size distribution. The probability distribution, $P_{cl}(s)$ is computed for a particular value of $r_{cl}^{(\pm \pm)}(=1.4\sigma)$. We plot $P_{cl}(s)$ for the $+ve$ charges for $r_{cl}^{(\pm \pm)}(=1.4\sigma)$ as a function of $s$ for different $f$ in Fig. 3(b). For $f=0$, $P_{cl}(s)$ shows high peak at $s=1$ indicating isolated clusters. The situation remains identical even for $f=50$ when the system is mostly dominated by small clusters. However, for both $f=150$ and $f=300$, the initial peak in $P_{cl}(s)$ decrease significantly while the probability increase for higher $s$. Mean cluster size of the system, $<s_{cl}>$ is given by $\int sP_{cl}(s)ds$ while average number of attached neighbors  $<\xi _{cl}> \sim \int (s-1)P_{cl}(s-1)ds$ excluding the particle in the reference. The spread in the distribution is given by $\sigma_{cl}^{2}=<\xi^{2}_{cl}>-<\xi _{cl}>^{2}$. In Inset Fig.2(b), with increasing $f$, $\sigma_{cl}^{2}$ show a maximum at $f=150$. 

The initial-homogeneous phase consists of tiny clusters while bigger ones span the system in the lane phase. In the intermediate phase, however, the distribution of domains is maximally broadened. The existing time-scales in the system are identified via the peaks in $\chi_{4}(t)$\cite{smk-cd}. But unlike glasses\cite{smk-cd}, here, the time scales are not separated by orders of magnitude although both show slowing down of dynamics. The intermediate heterogeneous phase involves mixed relaxation processes of both slow and fast particles in the system while we observe dominant response from increased proportion of slow particles in the lane phase. Interestingly, the shift of the peaks in $\chi^{(+)}_{4(\bot)}(t)$ with increasing $f$ follows the trend of the long-time-particle diffusion in the transverse plane: reports \cite{lane1,lane8} show there exist increased diffusion followed by diffusion with decreasing magnitude as the system approaches the laning transition. This in-plane-slowing-down was also probed in lane forming system via diffusion in 2-dimensions\cite{netz}. We now look for a generalized model that interprets some of these issues.

{\it Generalization:} Let us consider a situation: There are $N_{C}$ number of clusters with individual length-scales $s_{1}, s_{2}$ etc. that relax with different rates $\tau_{1}, \tau_{2} $ etc. The associated distribution is given by $P(s)$ and $\tilde{P} (\tau)$. Now we ask: what happens to particle diffusion if it propagates through a system of domains? Let the probability of displacement by $\Delta r$ in $\Delta t$ is $G_{S} (\Delta r, \Delta t)$. In presence of such domains in the system with distribution of length-scales, $P(s)$ and the associated time-scales of relaxations, using simple dimension analysis, one can write, $G_{S} (\Delta r, \Delta t) \sim \int G_{S}^{s|\tau} (\Delta r, \Delta t) P(s)\tilde{P}(\tau)  ds d\tau $ with $G_{S}^{s|\tau} (\Delta r, \Delta t)= \frac{\sqrt{\tau}}{s \sqrt{\Delta t}}  e^{-\frac{\tau\Delta r ^{2}}{s^{2} \Delta t}}$. For normal liquids,  $G_{S}^{s|\tau} (\Delta r, \Delta t) \rightarrow G_{S}^{(0)} (\Delta r, \Delta t)=\frac{1}{\sqrt{D\Delta t}} e^{-\frac{\Delta r ^{2}}{D\Delta t}}$ in the limits $s\rightarrow \infty$, $\tau\rightarrow \infty$ and $ \frac{s^2}{\tau}\rightarrow D$, the Diffusion coefficient. Hence, for our model system, using $\tilde{P(\tau})\sim \sum_{i=1}^{N_{C}} \delta(\tau -\tau_{i})$ and ${P(s})\sim \sum_{i=1}^{N_{C}} \delta(s -s_{i})$, we get $G_{S} (\Delta r, \Delta t) \sim \frac{1}{ \sqrt{\Delta t}}  [(\sum_{i=1}^{N_{C}} \frac{\sqrt{\tau_{i}}}{s_{i}}  e^{-\frac{\tau_{i} \Delta r ^{2}}{{s_{i}}^{2} \Delta t}})+ \frac{\sqrt{\tau_{1}}}{s_{0}}  e^{-\frac{\tau_{1} \Delta r ^{2}}{{s_{0}}^{2} \Delta t}} + \frac{\sqrt{\tau_{0}}}{s_{1}}  e^{-\frac{\tau_{0} \Delta r ^{2}}{{s_{1}}^{2} \Delta t}} +..]$ where cross terms are non-trivial and are entirely due to the domain interactions and hence, a heterogeneity in diffusion even in relatively high $\Delta t$ limit. This scenario was probed on the same system recently, via computer simulations \cite{lane10}.

We now qualitatively relate the heterogeneous response we find in our simulations to the dynamics of these diffusing clusters. Let us consider a specific overlap between configurations separated by time $t$ given by a unique time-scale of relaxation $\tau$: $Q(t,\tau)$. In presence of different domains with different relaxation times, the effective overlap is given by the weighted averaged value: $Q_{w}(t)\sim \int Q(t,\tau) \tilde{P} (\tau) d\tau$. For a system of such domains, $\tilde{P(\tau})\sim \sum_{i=1}^{N_{C}} \delta(\tau -\tau_{i})$. Thus, we obtain, $\chi_{4} (t)=<Q^{2}_{w}(t)>-<Q_{w}(t)>^{2} \sim [\sum_{i=1} ^{N_{C}} \chi^{(i)}_{4}(t)]+\chi^{(1,2)}_{4}(t)+..$, where $\chi^{(i)}_{4}(t)=<Q^{2}(t,\tau_{i})>-<Q(t,\tau_{i})>^{2}$, the response associated with $ith$ domain and $\chi^{(i,j)}_{4}(t)=<Q(t,\tau_{i})Q(t,\tau_{j})>-<Q(t,\tau_{i})><Q(t,\tau_{j})>$, the non-trivial correlation between the two coexisting relaxation processes of the domains $i$ and $j$. Thus, we obtain a heterogeneity in dynamical relaxation while the homogeneous limit is attained for $P(s)\sim N_{C} \delta(s-s_{0})$ and $\tilde{P(\tau})\sim N_{C} \delta(\tau -\tau_{0})$ as $s_{i}\rightarrow s_{0}$ and $\tau_{i}\rightarrow \tau_{0}$. 

Since $<\frac{s^{2}}{\tau}> \sim <D>$, $ \int \frac{s^{2}}{\tau}P(s)\tilde{P}(\tau) ds d\tau \sim \int D P'(D)dD$. Hence, one observes a broadening in $P'(D)$ for a broadened  $P(s)$ and $\tilde{P}(\tau)$ as shown in Ref. \cite{lane10}. Convoluting such a diffusion spectrum, $P'(D)$ over fundamental diffusion process, $G_{S}^{(0)} (\Delta r, \Delta t)$, one derives a heterogeneous $G_{S} (\Delta r, \Delta t) \sim \int G_{S}^{(0)} (\Delta r, \Delta t) P'(D) dD$ \cite{nat_mat} where the measure of heterogeneity lies in moments of $P'(D)$ \cite{lane10}. In contrast, in this model, this is given by $P(s)$ and $\tilde{P}(\tau)$ while the diffusivities ($\sim \frac{s^{2}}{\tau}$) are assumed to be local as in Ref. \cite{lane10}. Hence, in the limit $\tau_{i}\sim \tau_{0}+\Delta$ with $\Delta \rightarrow 0$, one finds $\chi_{4}(t)\sim [<Q^{2}(t,\tau_{0})>-<Q(t,\tau_{0})>^{2}]+ [< Q(t,\tau_{0})\delta Q>-<Q(t,\tau_{0})><\delta Q>]$   where $\delta Q \sim \Delta \frac{\partial Q}{\partial \tau}$ computed at $\tau=\tau_{0}$. The second term is responsible for broadening of the peak in $\chi_{4}(t)$ in presence of competing time scales in the system as seen for $f=150$ in our simulation data. For $f=300$, there is separation of time scales in bands due to presence of fast and slow particles in the system, hence, we find multiple peaks in $\chi_{4}(t)$.

{\it Discussion: }The power-law dependence in Overlap function has been found in the context of aggregate formers \cite{td} while the stretched exponential dependence is in agreement to the dynamic behavior of intermediate scattering function in the pre-lane phase \cite{lane10}. The non-monotonic dependence in cluster size distribution is likely to be related with the non-monotonicity of heterogeneity in diffusion reported in Ref. \cite{lane10}. There are attempts \cite{nat_mat,smk} in recent years to realize the heterogeneity in diffusion in term of a spectrum of diffusivities when the system dynamics is not entirely characterized by a unique diffusion co-efficient. Also there are Random Walk descriptions of the process \cite{ctrw}. In all these attempts, heterogeneous diffusion is realized in terms of the competing time-scales extracted by de-convoluting the associated cumulative distribution of particle displacements or the relaxation profiles\cite{pin}. This heterogeneity of diffusion has been linked to particle correlations\cite{lane10} as well when the particles diffuse in the neighborhood of a domain while this particular model interprets the heterogeneous relaxation in self-overlaps and heterogeneous response in dynamical susceptibility explicitly in terms of length-scales and time-scales of relaxing domains. However, this model is restricted to systems having $P(s)\sim \frac{1}{s^{\alpha}}$ with $\alpha >3$ or $P(s)$ having $\sim e^{-s}$ so that the average diffusivity, $ <D> \sim \int \frac{s^{2}}{\tau}P(s)\tilde{P}(\tau) ds d\tau$ remains always finite.

In conclusion, in a driven mixture of oppositely charged colloid, we probe competing relaxations of the lanes as the system approaches the laning transition from an initial homogeneous mixture. With increasing field, the lanes proliferate and their cumulative interactions grow monotonically. However, the cluster size distribution evolves non-monotonically. Since these individual length-scales relax differently, the system shows a heterogeneous mixed response in the intermediate phase while in the lane phase, there is a separation of time-scales in distinct peaks due to increased proportion of slow particles. With the increase of the field strength as the bigger lanes interact among themselves, the in-plane motion continues to slow down while they grow stronger. The heterogeneity appears when the size-distribution of these domains is maximally broadened. Unlike the role of temperature in glasses\cite{smk-cd}, the onset of this heterogeneity is primarily due to the competition between applied field and the particle interactions\cite{lane10}. This nontrivial heterogeneous response could be verified experimentally. Also, it would be interesting to check whether these rearrangements of domains affect the visco-elastic and dielectric response of the system, not only, in the present scenario but also in cases where similar charged or magnetic dipolar colloids are subject to oscillatory field, or confinement or both \cite{appl1,appl2}, in steady states, even in ageing conditions. Further, we show that the heterogeneity in dynamical responses is due to the heterogeneity in diffusion via a phenomenological model. Unlike using diffusion, the model uses only the time-scales and length-scales of these relaxing domains to characterize particle dynamics and subsequent heterogeneity in dynamical relaxation. Thus, the generality of our phenomenological model reinforces the possibility of the heterogeneity in diffusion not only in lane forming systems \cite{lane10}, but also in systems where such competing relaxations exist. On this note, we believe, this assay  opens up ranges of possibilities in unveiling unknown avenues.  

The author acknowledges J. Chakrabarti, T. Das and S. K. Paul for numerous insightful discussions, S. Bose and P. Tarafdar for critically reading the manuscript. 


\begin{thebibliography}{}

 

\bibitem{soft1}
 R. A. L. Jones,  Soft Condensed Matter, Oxford Master Series in Physics(Oxford University Press, Oxford, 2002)

\bibitem{soft2}
D. David Andelman, and G. Reiter, (Ed) Series in Soft Condensed Matter, Vol-1-6, (World Scientific, Singapore, 2012).

\bibitem{rev1}
 H. L{\" o}wen, Phys.\ Rep.\ , {\bf 237}, 249 (1994); H. L{\" o}wen,  J.\ Phys.\: Condens.\ Matter\ , {\bf 13}, R415(2001)
 
 \bibitem{rev2}
 A. V. Blaaderen  {\it et al.},  Eur.\ Phys.\ J.\ Special\ Topics, {\bf 222,} 2895 (2013); H. L{\"o}wen, Eur.\ Phys.\ J.\ Special\ Topics, {\bf 222}, 2727 (2013)

\bibitem{lane1}
J. Dzubiella, G. P. Hoffmann and H. L{\" o}wen, Phys. Rev. E, {\bf 65}, 021402 (2002)


\bibitem{lane2}
J. Chakrabarti, J. Dzubiella, H. L{\" o}wen,  Europhys.\ Lett.\ , {\bf 61}, 415 (2003)

\bibitem{lane3}
J. Chakrabarti, J. Dzubiella, and H. L{\" o}wen,  Phys.\ Rev.\ E, {\bf 70}, 012401 (2004)

\bibitem{netz}
R. R. Netz,  Europhys.\ Lett.\ , {\bf 63}, 616 (2003)

\bibitem{lane3a}
H. L{\" o}wen and J. Dzubiella, Faraday Discuss., {\bf 123}, 99 (2003)

\bibitem{lane4}
M. E. Leunissen  {\it et al.}, Nature, {\bf 437}, 235 (2005)

\bibitem{lane5}
M. Rex and  H. L{\" o}wen,  Phys.\ Rev.\ E,  {\bf 75}, 051402 (2007) 

\bibitem{lane6}
K. R. S{\"u}tterlin, {\it et al.},  Phys.\ Rev.\ Lett., {\bf 102}, 085003 (2009)

\bibitem{lane7}
T. Vissers, A. van Blaaderen, and A. Imhof, Phys. Rev. Lett., {\bf 106}, 228303 (2010)

\bibitem{lane8}
T. Vissers, {\it et al.}, Soft Matter, {\bf 7}, 2352 (2011)


\bibitem{lane9}
T. Glanz, and H. L{\"o}wen, J.\ Phys.:\ Condens.\ Matter\ , {\bf 24}, 464114 (2012)

\bibitem{lane10}
S. Dutta and J. Chakrabarti (To be published); S. Dutta and J. Chakrabarti, arXiv:1607.08341

\bibitem{lane11}
K. Klymko, P. L. Geissler and S. Whitelam, Phys.\ Rev.\ E, {\bf 94}, 022608 (2016)

 \bibitem{ant}
 I. D. Couzin and N. R. Franks, Proc. R. Soc. London, Ser. B,  {\bf 270}, 139 (2003)

 \bibitem{ped}
D. Helbing, L. Buzna, A. Johansson and T. Werner, Transport. Sci., {\bf 39}, 1 (2005)

\bibitem{granule}
M. P. Ciamarra, A. Coniglio and M. Nicodemi, J. Phys.: Condens. Matter, 17, S2549 (2005)

\bibitem{plasma}
K. R. S{\"u}tterlin {\it et al.} , Phys. Rev. Lett., {\bf 102}, 085003 (2009)

\bibitem{swimmers}
F. Kogler and S. H. L. Klapp, Europhys.\ Lett.\ , {\bf 110}, 10004 (2015)

\bibitem{smk}
S. Sengupta, and S. Karmakar, J.\ Chem.\ Phys., {\bf 140}, 224505 (2014)

\bibitem{nat_mat}
B. Wang {\it et al.}, Nat.\ Mat., {\bf 11}, 481 (2012).

\bibitem{appl1}
N. Cevheri and M. Yoda, Lab Chip {\bf 14}, 1391 (2014)

\bibitem{appl2}
I. S. Aranson, Phys. \ -Usp.\ , {\bf 56}, 79 (2013)
 
\bibitem{erm}
D. L. Ermak,  J.\ Chem.\ Phys. , {\bf 62}, 4189 (1975)


\bibitem{cd}
C. Dasgupta, A. V. Indrani, S. Ramaswamy and M. K. Phani, Europhys. Lett., {\bf 15}, 307 (1991)


\bibitem{smk-cd}
S. Karmakar, S. Sastry and C. Dasgupta, Proc. Natl. Acad. Sci. USA, {\bf 106}, 3675 (2009), S. Karmakar, C. Dasgupta and S. Sastry, Annu.\ Rev.\ Condens.\ Matter.\ Phys {\bf 5}, 255 (2014)

\bibitem{td}
T. Das, T. Lookman, and M. M. Bandi, arXiv:1505.05702 (2015)


\bibitem{allen}
M. P. Allen and D. J. Tildesley, Computer Simulation of Liquids (Oxford Science Publications, Oxford, 1989) 

\bibitem{ef_int}
J-P. Hansen and I. R. McDonald, Theory of Simple Liquids (Academic Press, 2006)


\bibitem{ctrw}
M. V. Chubynsky and G. W. Slater, Phys.\ Rev.\ Lett.\ ,{\bf 113}, 098302 (2014); A. G. Cherstvy and R. Metzler, Phys.\ Chem.\ Chem.\ Phys.\ DOI: 10.1039/C6CP03101C (2016)

\bibitem{pin}
B. P. Bhowmik, R. Das and S. Karmakar, J.\ Stat.\ Mech.\ 074003 (2016)

\end{thebibliography}

\end{document}